\documentclass[english,twocolumn]{article}

\usepackage[utf8]{inputenc}
\usepackage{amsmath,amssymb,bm}
\usepackage{braket}
\usepackage{physics}

\usepackage{graphicx}
\usepackage{float}

\usepackage{tabularx}
\usepackage{multirow}
\usepackage{dcolumn}

\usepackage{geometry}
\usepackage{authblk}

\usepackage{multicol}

\usepackage[dvipsnames]{xcolor}
\usepackage{soul} 

\usepackage{afterpage}
\usepackage{subfiles}
\usepackage{lipsum}
\usepackage{bbold} 

\usepackage[square,numbers,sort&compress]{natbib}
\usepackage{hyperref}

\newcommand{\la}{\left\langle}
\newcommand{\ra}{\right\rangle}
\newcommand{\ls}{\left[}
\newcommand{\rs}{\right]}
\newcommand{\lp}{\left(}
\newcommand{\rp}{\right)}
\newcommand{\lbr}{\left\{}
\newcommand{\rbr}{\right\}}

\begin{document}

\title{Machine Intelligence on Wireless Edge Networks}
\date{}
\author[1,*]{Sri Krishna Vadlamani}
\author[1,*]{Kfir Sulimany}
\author[2]{Zhihui Gao}
\author[2]{Tingjun Chen}
\author[1]{Dirk Englund}

\affil[1]{Research Laboratory of Electronics, Massachusetts Institute of Technology, Cambridge, MA, USA}
\affil[2]{Department of Electrical and Computer Engineering, Duke University, Durham, NC 27708, USA}
\affil[*]{These authors contributed equally to this work.}

\twocolumn[\begin{@twocolumnfalse}

\maketitle
\vspace{-3em}
\begin{abstract}
Machine intelligence on edge devices enables low-latency processing and improved privacy, but is often limited by the energy and delay of moving and converting data. Current systems frequently avoid local model storage by sending queries to a server, incurring uplink cost, network latency, and privacy risk. We present the opposite approach: broadcasting model weights to clients that perform inference locally using in-physics computation inside the radio receive chain. A base station transmits weights as radio frequency (RF) waveforms; the client encodes activations onto the waveform and computes the result using existing mixer and filter stages—RF components already present in billions of edge devices such as cellphones—eliminating repeated signal conversions and extra hardware. Analysis shows that thermal noise and nonlinearity create an optimal energy window for accurate analog inner products. Hardware-tailored training through a differentiable RF chain preserves accuracy within this regime. Circuit-informed simulations, consistent with a companion experiment, demonstrate reduced memory and conversion overhead while maintaining high accuracy in realistic wireless edge scenarios.
\newline
\end{abstract}
\end{@twocolumnfalse}]

\section{Introduction}

Machine intelligence on edge devices enables perception, control, and decision-making close to where data is generated. This reduces latency, saves network bandwidth, and improves privacy for applications ranging from wearables and augmented reality to industrial monitoring and autonomous systems. However, running modern machine-learning models on the edge is often constrained not by computational throughput, but by the cost of moving and converting data. Transporting parameters and activations between memory and processing units, and converting signals between analog and digital domains, dominate both energy consumption and latency \cite{sze2017efficient,horowitz20141,jouppi2017datacenter,satyanarayanan2017emergence,yang2017method}.

A common way to avoid storing large models locally is to send sensor data or features to a remote server for inference. This offloading paradigm removes the need for local model storage but consumes uplink bandwidth, increases energy use, adds network latency, and can compromise privacy \cite{vepakomma2018split,li2017multi,sulimany2024quantum}. These drawbacks are compounded when many devices share backhaul resources or operate under unreliable connectivity.

In this work we explore the opposite approach: instead of sending data to the server, we send the model to the client. Model parameters can be broadcast to many clients, amortizing their transmission cost over time and across devices, while allowing raw sensor data to remain local. The challenge is enabling such an architecture without requiring the client to store the full model or to repeatedly convert signals between analog and digital form between layers.

A promising direction is in-physics computation, in which core operations of inference or learning are performed directly within existing physical signal paths \cite{Momeni2025, momeni2023backpropagation, wright2022deep,wetzstein2020inference, sebastian2020memory}, eliminating the need for separate digital compute and reducing data motion. This principle has been demonstrated in photonic systems that leverage light for high-throughput, energy efficient, low-latency, or quantum-secure computations \cite{shastri2021photonics,xue2024fully,yildirim2024nonlinear,mcmahon2023physics,ashtiani2022chip,feldmann2021parallel,chen2023deep,wang2022optical,xu202111,bogaerts2020programmable, sulimany2024quantum,bandyopadhyay2025three, bacvanski2024qamnet,miller2017attojoule, berloff2025analog,xu2024large,lin2018all}. Related concepts appear in processing-in-memory platforms, where vector–matrix products and logic are computed directly in memory arrays, including crossbar and three-dimensional architectures \cite{shafiee2016isaac,chi2016prime,ankit2019puma,huo2022computing,rasch2024fast,lammie2025inherent,pedretti2021memory,jung2022crossbar,cheon20232941,tagata2024double,cai2020power, le202364}. Logic-in-memory approaches and low-precision learning further align computation with device operating ranges \cite{leitersdorf2021multpim,leitersdorf2022matpim,kvatinsky2014magic,hubara2018quantized,hubara2016binarized,jebali2024powering}.

In wireless networks, the natural superposition of signals over the multiple-access channel has long been exploited for over-the-air computation, allowing functions such as sums to be evaluated directly in the medium rather than after demodulation \cite{nazer2007computation,goldenbaum2013robust,goldenbaum2014nomographic,csahin2023survey}. Recent work has shaped propagation or modulation to implement neural network primitives, including reconfigurable-surface-assisted layers and wireless mappings of fully connected layers \cite{sanchez2022airnn,reus2023airfc,choi2024wireless,stylianopoulos2025universal,razavikia2023computing}. RF circuit research has demonstrated microwave-domain neural operations, from reconfigurable analog matrix multipliers to integrated microwave neural networks and continuous-time reservoirs \cite{zhu2023reconfigurable,govind2025integrated,senanian2024microwave}. These methods reduce communication or implement analog primitives, but they generally rely on engineered propagation environments or specialized front ends, and do not provide end-to-end inference on unmodified commodity clients without local weight storage for deployment in billions of existing wireless edge devices.

\begin{figure*}[t]
    \centering
\includegraphics[width=\textwidth]{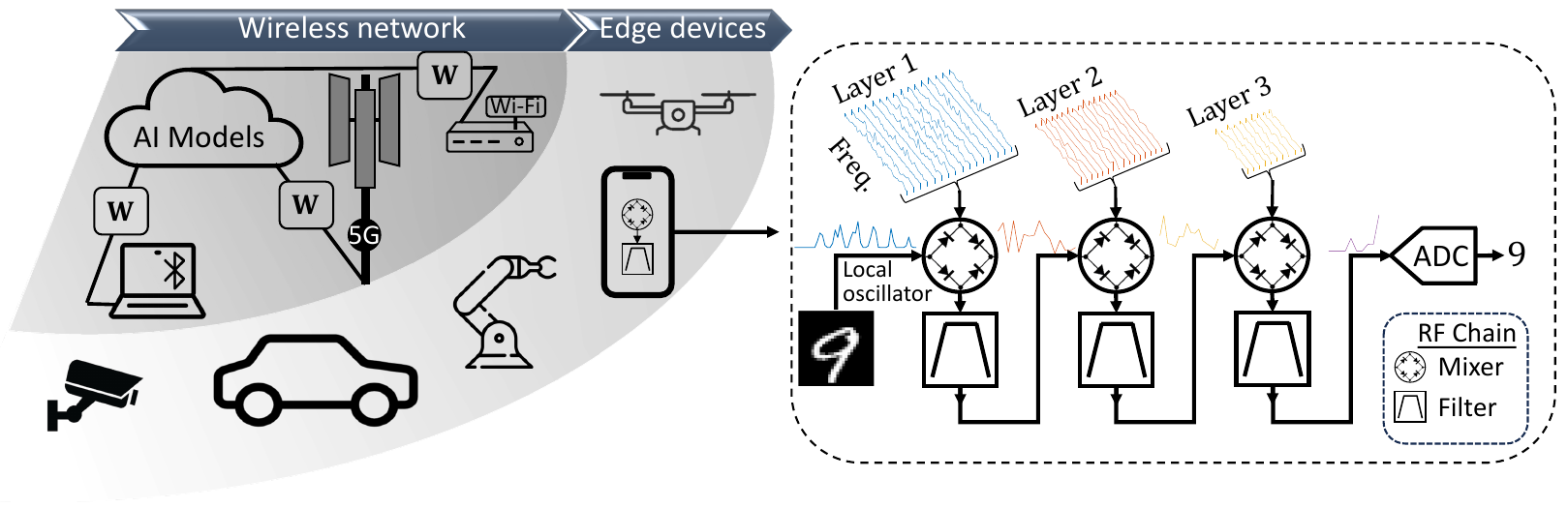}
     \caption{\textbf{System illustration.} The edge device utilizes the existing radio frequency (RF) analog chain of wireless transceivers to perform disaggregated deep learning. The weights, denoted as $\mathbf{w}$, are transmitted as RF signals from a server (left), encoded in either the frequency or time domain. These signals are received by the edge device, where they are multiplied by the data signals using an RF diode ring mixer. The signal is then filtered and passed through the mixer again for the next multiplication (right). In the top right, we present the frequency-encoded signals (see main text) representing the numeric weights of a three-layer model trained to classify the MNIST dataset. The weights are not stored on the client, which mixes them in-path with locally encoded activations and samples the result using an analog-to-digital converter (ADC).}
 \label{scheme}
 \end{figure*}

Here, we apply in-physics computation to wireless edge inference by placing the core multiply–accumulate operations inside the client’s radio receive path (Fig.~\ref{scheme}). A base station broadcasts the model weights as RF waveforms; the client encodes activations onto a carrier and uses the standard mixer and filter stages of its receive chain to perform the necessary multiplications and accumulations \cite{gilbert1968precise,sulivan1997low,ZEM-4300+}. This removes the need to store the model locally and avoids repeated signal conversions between layers. Unlike array-based analog accelerators, our design uses the RF hardware already present in commodity transceivers; unlike channel-engineered schemes, it requires no reconfigurable surfaces or controlled propagation. The result decouples model capacity from client memory and shifts cost from data movement to RF front-end operations already amortized by communication \cite{bergman2022reconsidering}.

We analyze the accuracy limits of this architecture, showing how thermal noise and front-end nonlinearity produce an energy–accuracy trade-off with a well-defined optimal operating range. This behavior is consistent with classical thermal-noise theory for receivers \cite{nyquist1928thermal,callen1951irreversibility} and aligns with a companion experimental demonstration of disaggregated deep learning at radio frequencies with ultra-low energy consumption \cite{gao2025disaggregated}. We also introduce two end-to-end pipelines: a digital-tailored version that merges Fourier transforms into the weights via IQ modulation \cite{abari201427,bloessl2013ieee}, and a hardware-tailored version that trains through a differentiable mixer–filter model to maintain accuracy without intermediate signal conversions \cite{wright2022deep,xue2024fully,pai2023experimentally,shastri2021photonics}. We refer to this family of architectures as \emph{Machine Intelligence on Wireless Edge Networks (MIWEN)}.

\section{Results}

We first show how standard wireless-receiver mixers enable matrix–vector multiplication (MVM) using time- and frequency-multiplexed encodings (Sec.~\ref{sec2.1}), as illustrated in Fig.~\ref{scheme}. We then analyze accuracy limits imposed by thermal noise and mixer nonlinearity
and translate them into an effective‑bit metric and energy‑per‑MAC trade‑offs (Sec.~\ref{sec2.2}). Building on these primitives, we present two computation pipelines: a digital-tailored variant that absorbs the receiver’s Fourier transform stage into the model weights via IQ modulation and demodulation, eliminating the need for explicit Fourier transform at the client, and a hardware‑tailored variant that trains through a differentiable model of the mixer–filter chain to avoid inter‑layer conversion (Sec.~\ref{sec2.3}). Finally, we report classification accuracy results versus total input energy.

\subsection{Matrix-Vector Multiplication via Wireless Receivers}\label{sec2.1}
Modern deep learning hardware systems require the ability to efficiently implement matrix-vector multiplication (MVM). Inspection of the block diagrams of wireless receivers reveals that they contain an analog frequency mixer that performs frequency up- and down-conversion in the standard communication pipeline. An ideal mixer takes in waveforms $w(t)$ and $x(t)$, and outputs the instantaneous product waveform $y(t)=w(t) \cdot x(t)$~\cite{gilbert1968precise}. Receiver mixers can be utilized in different ways to calculate MVM, depending on how the matrix and vector are encoded into the RF signals. 

\subsubsection{Time-Multiplexed Encoding}

One way of using a mixer to calculate the inner product of vectors $\mathbf{w},\mathbf{x}\in\mathbb{R}^d$, given by $\mathbf{w}\cdot \mathbf{x}$, is to encode the components of each vector into the amplitudes of two separate time-multiplexed voltage trains of $d$ pulses each, $w(t)$ and $x(t)$, and send them as inputs to the mixer.
The output $y(t)$ is then a voltage train of $d$ pulses whose amplitudes are element-wise products of the vectors $\mathbf{w}$ and $\mathbf{x}$. An op-amp integrator can be used to integrate this pulse train to obtain the final inner product $\mathbf{w\cdot x}$. To generalize this to the matrix-vector product of matrix $\mathbf{W}\in\mathbb{R}^{m\times d}$ and vector $\mathbf{x}\in\mathbb{R}^d$, one could make $m$ identical copies, denoted by $\mathbf{x}^{(i)}$, of the vector $\mathbf{x}$ (or fanout) and compute for each index $i$ the inner product of $\mathbf{x}^{(i)}$ with $\mathbf{w}^{(i)}$, the $i$-th row of $\mathbf{W}$, hence yielding all the components $y_i$ of $\mathbf{y}=\mathbf{W} \cdot \mathbf{x}$. This method was proposed and investigated in~\cite{hamerly2019large}. While this approach is simple and appealing, wireless receivers rarely contain integrating elements. Therefore, to obtain a practical implementation, we would need to convert the output product pulse trains $y^{(i)}(t)$ to the digital domain before summing up the pulse amplitudes to obtain the inner products $y_i$ that make up the output vector $\mathbf{y}$. The analog-to-digital conversion erases all the energy and latency gains from the previous analog mixer multiplication step.

\subsubsection{Frequency‑Multiplexed Encoding}

In frequency-encoding, the weights $\lbr W_{pq} | p\in[m],q\in[d]\rbr$ (where we use $[n]$ to denote the set $\lbr 1, \dots, n \rbr$) of a matrix $\mathbf{W} \in \mathbb{R}^{m\times d}$ are encoded into the amplitudes of the frequency components of a single waveform $w(t)$, while the components $\lbr x_{r}|r\in[d]\rbr$ of the vector $\mathbf{x}$ are encoded into the amplitudes of the frequency components of a different waveform $x(t)$. More precisely, we set:
\begin{align}
    w(t)&=\lp\sum_{r=1}^m\sum_{n=1}^dW_{rn}e^{i((r_0+r)\Delta\omega_s+(n_0+n)\Delta\omega_l)t}\rp e^{i\omega_at}+\mathrm{c.c.}\label{eq:wt}\\
    x(t)&=\lp\sum_{p=1}^dx_pe^{i((n_0+p)\Delta\omega_l)t}\rp e^{i\omega_bt}+\mathrm{c.c.}\label{eq:xt},
\end{align}
where $\omega_a,\omega_b$ are the weight and activation waveform carrier frequencies, $\Delta\omega_l$ is a larger `super-frequency spacing', $\Delta\omega_s$ is a smaller `sub-frequency spacing', and c.c refers to the complex conjugate of the preceding terms that is added in to ensure that the resulting signal is real-valued. 

The basic idea of the above `comb'-like encoding is that all the elements of the first column of $\mathbf{W}$ are modulated into the amplitudes of consecutive frequency comb lines spaced by $\Delta\omega_s$ intervals, followed by all the elements of the second column of $\mathbf{W}$, and so on. $\Delta\omega_l$ is the spacing between the comb lines that encode the column leaders, $W_{11}, W_{12}, \dots$. The column leader spacing $\Delta\omega_l$ has to be chosen to be equal to $m\Delta\omega_s$ so that all elements of the $q$-th column of $\mathbf{W}$ are accommodated between the $q$-th column leader and the $(q+1)$-th column leader comb lines. We emphasize once again that all elements of the matrix $\mathbf{W}$, as opposed to just a single column, are encoded into the waveform in this method. The encoding of the vector $\mathbf{x}$ is more straightforward, with the components being encoded onto the amplitudes of comb lines spaced $\Delta\omega_l$ apart.    

This method has been used in the optical domain to demonstrate ultra-efficient MVM with minimal analog-to-digital conversions \cite{davis2022frequency}. We shall, however, focus on achieving multiplication through the native RF mixers in wireless receivers. The output waveform $y(t)$ is:
\begin{align}
y(t) &= w(t)\,x(t) \\
&= e^{\,i(\omega_a+\omega_b)t}\,Y_{+}(t)
   + e^{\,i(\omega_a-\omega_b)t}\,Y_{-}(t)
   + \mathrm{c.c.} \notag
\end{align}

\begin{align*}
Y_{+}(t) &= \sum_{r=1}^{m} \sum_{n,p=1}^{d}
            W_{rn}\,x_{p}\;
            e^{\,i\{(r_{0}+r)\Delta\omega_{s}+(2n_{0}+n+p)\Delta\omega_{l}\}t},\\[2pt]
Y_{-}(t) &= \sum_{r=1}^{m} \sum_{n,p=1}^{d}
            W_{rn}\,x_{p}^{*}\;
            e^{\,i\{(r_{0}+r)\Delta\omega_{s}+(n-p)\Delta\omega_{l}\}t}.
\end{align*}
From the preceding expression, one sees that the output $y(t)$ is itself a frequency comb that is centered at the sum and difference carrier frequencies, $\omega_a+\omega_b$ and $\omega_a-\omega_b$. It can be seen that the output comb line at $\omega_r = \omega_a - \omega_b + (r_0+r)\Delta\omega_s$ has the amplitude:
\begin{equation}\label{eq:matvectprod}
y(\omega_r)=\sum_{n=1}^{d} W_{rn}\,x_n^{*},
\end{equation}
which is the $r$-th component of $\mathbf{W} \cdot \mathbf{x}^*$. One could therefore capture only the matrix-vector product by using a bandpass filter to pass all the components of $\mathbf{W} \cdot \mathbf{x}^*$ while blocking all other spurious frequencies. We shall use this kind of filtration repeatedly in the paper.

Since the frequency-encoding method achieves full MVM through a single use of the mixer without the need for an integrator, it is naturally suited for implementation on standard wireless receivers. Our concrete proposal is to implement this technique on RF receivers to achieve energy- and space-efficient machine learning inference on size, weight, and power (SWaP)‑constrained edge devices and sensors. In this protocol, a base station streams out the weights $\mathbf{W}$ of a model, in the form of the frequency-multiplexed signal $w(t)$, over a wireless channel which are then received and applied by wireless clients on their local activations $\mathbf{x}$, which are themselves pre-encoded into the frequency-multiplexed signal $x(t)$. 

\subsection{Fundamental Limits on Physical Inner Product Computations} 
\label{sec2.2}
Since physical systems operate at finite temperatures, physical computations naturally face interference from thermal noise, caused by fluctuating energy exchanges between the system and its surroundings.
In an analog inner product engine for instance, this noise mechanism leads to discrepancies between the physically computed inner product and the true inner product. Discrepancies are also caused by the imperfect nature of analog multipliers. We go into more details on each of these concepts next.

\subsubsection{Noise, Mixer Non‑Idealities, and Effective Precision}
 \paragraph{Noise}
Every resistor in an electrical circuit, by virtue of the fluctuation-dissipation theorem, produces fluctuating thermal noise voltages, referred to as Johnson-Nyquist noise, across its terminals \cite{nyquist1928thermal}. The Johnson noise voltage $\Delta v^{(\text{th})}(t)$ across a resistor $R$ at time $t$ is a random white noise with mean $0$ and variance $4kTR\Delta f$ where $\Delta f$ is the bandwidth of frequencies allowed by the circuit, $k$ is the Boltzmann constant, and $T$ is the temperature of the resistor. 

 \paragraph{Mixing in RF circuits}

Active mixers called Gilbert cells \cite{gilbert1968precise} can perform exact multiplication of two RF waveforms $w(t)$ and $x(t)$ to produce the output $y(t) = w(t) x(t)$. The downside is that they are composed of transistors and need to be driven actively. Passive mixers, such as the diode ring mixer, on the other hand, consume no energy to be operated but tend to perform only approximate multiplication. For weight and activation signals $w(t),x(t)$, the diode ring mixer operates as follows:
\begin{align}
y(t)=\frac{w(t)}{2}+\frac{V_T}{2} \cdot \ln{\lp\frac{e^{\frac{x(t)}{V_T}}+e^{\frac{-w(t)}{V_T}}}{e^{\frac{x(t)}{V_T}}+e^{\frac{w(t)}{V_T}}}\rp},
\label{eq:dioderingeq}
\end{align}
in which $V_T=kT/e$, where $k,T$ have their usual meanings and $e$ is the electron charge. The derivation of this formula is provided in the Methods \ref{Met_cir}. At small signal levels the mixer behaves as an effective multiplier with gain $\approx 1/(4V_T)$ (Methods \ref{Met_cir}), which explains the initial accuracy increase before high‑order terms dominate at larger powers.

We assume that the two input ports of the mixer have input impedances of $R_w$ and $R_x$, which add Johnson noise $\Delta v^{(\text{th})}_w(t)$ and $\Delta v^{(\text{th})}_x(t)$ to their respective input waveforms. Moreover, we assume that the mixer's output impedance $R_y$ adds Johnson noise $\Delta v^{(\text{th})}_y(t)$ to the mixer's output waveform. The electronic shot noise in the system is ignored. Therefore, the final model for the mixer is:
\begin{align}
    y(t)=&\frac{w(t)+\Delta v^{(\text{th})}_w(t)}{2}\label{eq:noisydiodering}\\
    &+\frac{V_T}{2} \cdot \ln{\lp\frac{e^{\frac{x(t)+\Delta v^{(\text{th})}_x(t)}{V_T}}+e^{\frac{-w(t)-\Delta v^{(\text{th})}_w(t)}{V_T}}}{e^{\frac{x(t)+\Delta v^{(\text{th})}_x(t)}{V_T}}+e^{\frac{w(t)+\Delta v^{(\text{th})}_w(t)}{V_T}}}\rp}\notag
\end{align}

 \paragraph{Information content of physical inner products}

\textbf{Mutual information}: A common metric to measure the discrepancy between digital and analog computation is to define the average relative absolute error $E_{\text{rel}}~=~\langle|(\mathbf{w}~\cdot~\mathbf{x})_{\text{analog}}~-~(\mathbf{w}~\cdot~\mathbf{x})_{\text{digital}}|/|(\mathbf{w} \cdot \mathbf{x})_{\text{digital}}|\rangle_{\mathbf{w},\mathbf{x}}$, where the angular bracket notation $\langle\cdot\rangle$ denotes averages and the subscript indicates that the averaging is performed over pairs of vectors $\mathbf{w},\mathbf{x}$ sampled from some chosen distribution.

A shortcoming of this metric is that it could fail to meaningfully connect to the actual information theoretic transformations caused during the computation. A more appealing and theoretically robust metric is to compute the amount of information about the exact computation that is preserved by the physical machine in the presence of analog noise. This idea is captured exactly by the mutual information $I((\mathbf{w} \cdot \mathbf{x})_{\text{analog}}; (\mathbf{w} \cdot \mathbf{x})_{\text{digital}})$ between the analog and digital inner products. 

\textbf{ENOB: Effective number of bits}: Another way of evaluating the performance is through the signal-to-noise ratio of the output. Let a general noisy inner product engine be given by:
\begin{align}
    y=\left[\sum_i\left( w_i+n^{(w)}_i\right)\left( x_i+n^{(x)}_i\right)\right]+n^{(y)}.
\end{align}
The squared average output $\langle y\rangle_n^2$ for a given $\mathbf{w},\mathbf{x}$, but averaged over all the noise components, is $\langle y\rangle_n^2=\left(\sum_iw_ix_i\right)^2=\left(\mathbf w\cdot\mathbf x\right)^2$. The ``noise strength" $\text{Var}_n(y)$ for the same $\mathbf{w},\mathbf{x}$ is given by the variance of $y$ computed over the noise random variables. The signal and noise strengths for the overall setup are defined as the average of $\langle y\rangle_n^2$ and $\text{Var}_n(y)$ over all choices of $\mathbf{w},\mathbf{x}$: $S=\langle\langle y\rangle_n^2\rangle_{\mathbf{w},\mathbf{x}},\ N=\langle\text{Var}_n(y)\rangle_{\mathbf{w},\mathbf{x}}$. The SNR of our inner product engine is then $\text{SNR}=\frac{S}{N}$. One can then define the effective number of bits ENOB in the output as:
\begin{align}
    \text{ENOB}=\frac{1}{2}\log_2{\left(1+\text{SNR}\right)}\label{eq:ENOB}
\end{align}
We use this ENOB as an information‑theoretic precision metric for all subsequent plots and ablations. It turns out that this definition of ENOB can in fact be derived from the mutual information $I((\mathbf{w} \cdot~\mathbf{x})_{\text{analog}}; (\mathbf{w} \cdot~\mathbf{x})_{\text{digital}})$ introduced earlier \textemdash\ Methods~\ref{met_4.1} provides this derivation. We will use this definition to compute the information theoretic bit precision of our inner product engine in the next section.

\begin{figure}[t]
    \centering
\includegraphics[width=\columnwidth]{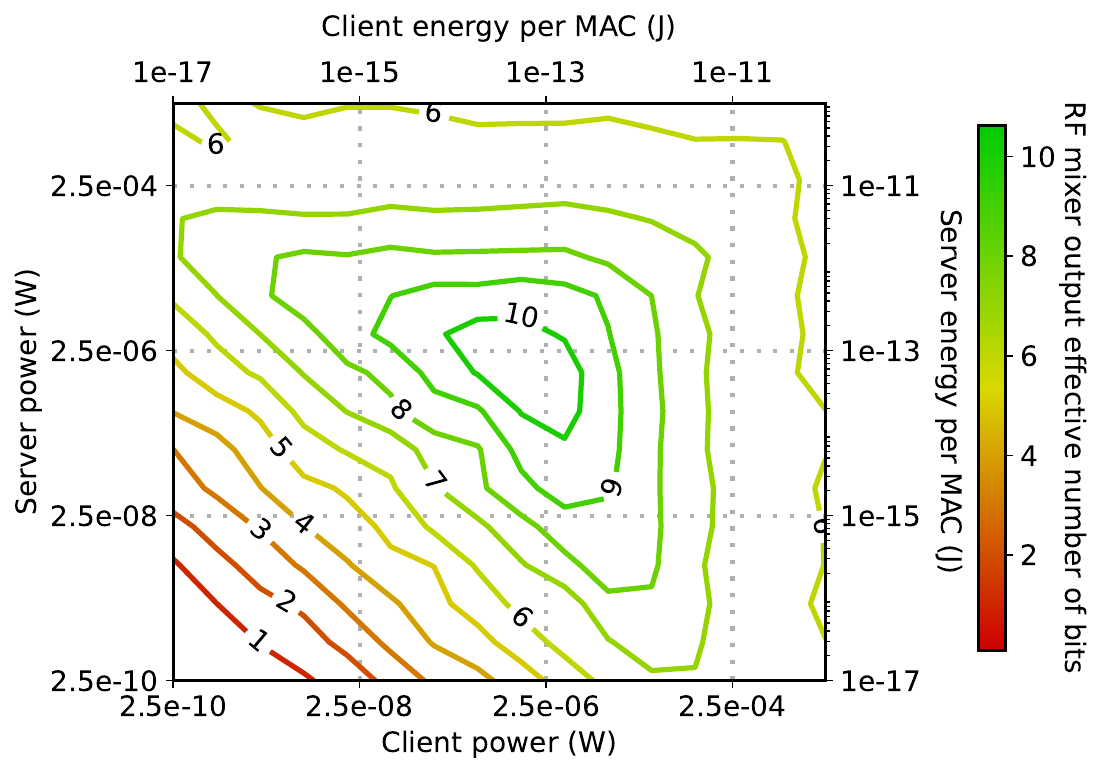}
     \caption{ \textbf{Effective number of bits (ENOB) for inner product computation vs power consumption.} ENOB in the output of the radio frequency inner product computation $\mathbf{w}\cdot\mathbf{x}$ is plotted as a function of client and server energy per Multiply-Accumulate (MAC) for a diode ring mixer. Here “client energy per MAC” accounts for the client’s activation waveform energy at a $50\,\Omega$ input; “server energy per MAC” accounts for the transmitted weight waveform energy; ADC energy is excluded in both.
     Contour labels denote ENOB. Axes show client/server power (W, log scale); top axes show corresponding per-MAC energies (J) A fixed bandwidth of 25 MHz is used to encode the components of the input vectors $\mathbf{w},\mathbf{x}$, the vector length is 256, ENOB is computed over 200 joint random initializations of the vectors $\mathbf{w},\mathbf{x}$. The inner product accuracy improves with increasing client and server power, reaches a peak, and reduces with further increases in input power due to the deviation of the diode ring transfer function from exact multiplication; see the main text for more details.  }
 \label{fig:enob}
 \end{figure}

\subsubsection{RF Energy per MAC for Inner‑Product Computation}
In this section, we present results for the variation of the bit precision of the inner products produced by a diode ring mixer as a function of the power of the weights and activation signals input to it. To this end, we developed custom layers in PyTorch for each step in the computation pipeline of the RF inner product scheme. The weight and activation vectors $\mathbf{w}, \mathbf{x}\in\mathbb{R}^d$ are sampled independently from a unit normal distribution and encoded into voltage waveforms $w(t),x(t)$ as described in the frequency-encoding section earlier. The waveforms $w(t),x(t)$ are then inserted into Eq.~\eqref{eq:dioderingeq} and the resultant inner product (extracted from the appropriate comb line in the frequency domain) is then plugged into Eq.~\eqref{eq:ENOB} to obtain the effective number of bits of the calculation. The weights or activations power incident on the mixer input ports is computed using the standard $V^2/R$ formula where $R$ here is assumed to be an input resistance of 50 $\Omega$ (see Methods \ref{app:SNR} for detailed definitions of power, energy, and energy per MAC). 

Fig.~\ref{fig:enob} presents the bit precision of the inner product computation (ENOB) as a function of the energy expenditure incurred by the client and the server. In the bottom-left corner of the figure, the accuracy of the inner product improves with increasing client or server power. The accuracy reaches a peak at a certain client power and server power and then deteriorates with further increases in the energy of either input. This is due to the mathematical behavior of the diode ring mixer: at low input powers, its transfer function Eq.~\eqref{eq:dioderingeq} is well approximated by the lowest Taylor series terms in the voltages $w(t),x(t)$ and behaves like an exact multiplier $w(t)x(t)$. For higher input powers, the higher Taylor series terms become important and the overall transfer function deviates from exact multiplication, leading to a drop in inner product accuracy.

\subsection{Machine Learning Performance} \label{sec2.3}
In this section, we discuss the extension of the frequency-domain inner product protocol to the case of multi-layer neural networks. We will discuss two versions: a digital-tailored protocol that performs conventional matrix-vector products and utilizes intermediary analog-to-digital and digital-to-analog conversions, and a different, hardware-tailored protocol that more naturally utilizes the physics of the RF pipeline itself. 

\begin{figure*}[t]
\begin{centering}
\includegraphics[width=\textwidth]{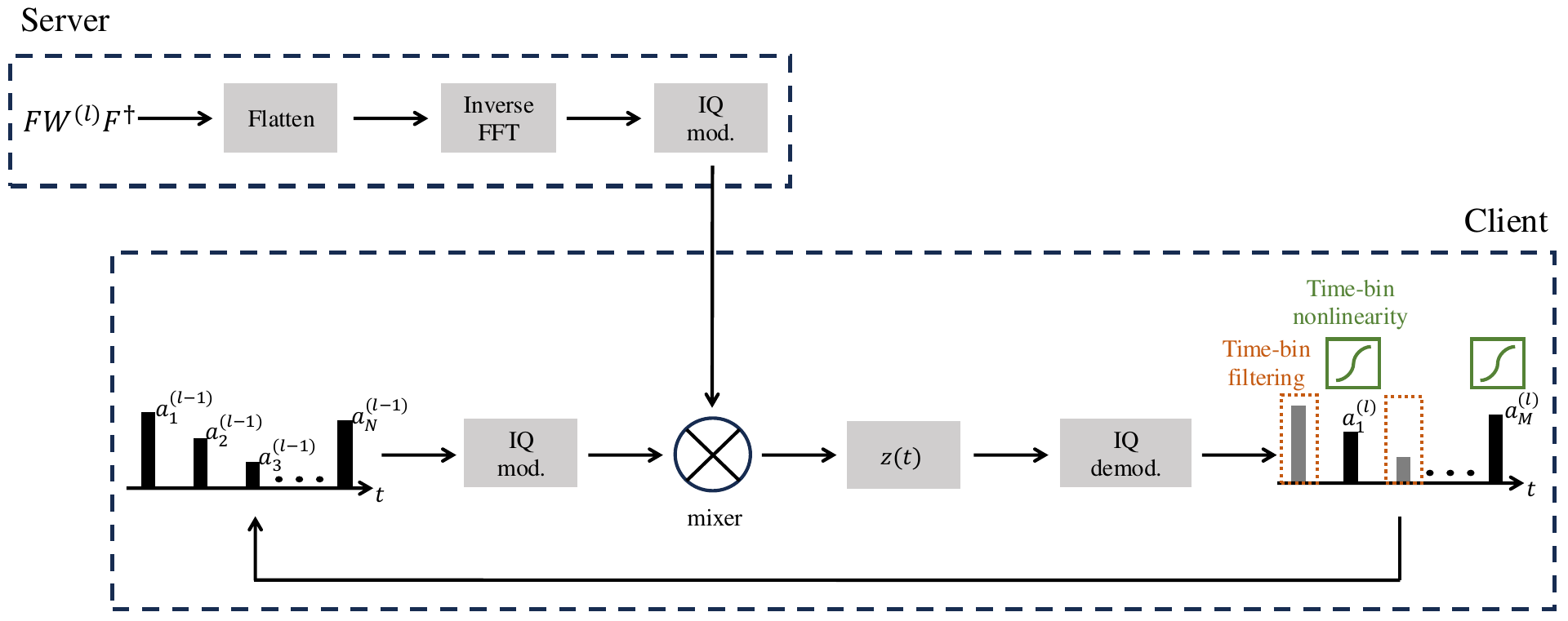}
\par\end{centering}
     \caption{\textbf{Digital-tailored machine learning computation pipeline.}
    In our digital-tailored proposal, the client only makes use of IQ modulators and demodulators and completely eschews the digital FFT and IFFT operations. Spurious frequency produced by the multiplication process appear as spurious time bins after IQ demodulation and are temporally filtered out. The activations are passed through a time-bin nonlinearity before being looped around to get multiplied by the incoming weights of the next layer from the server.}
 \label{fig:digitaltailored}
 \end{figure*}

In the digital-tailored approach (Fig.~\ref{fig:digitaltailored}), the neural network inputs $\mathbf{x}$ and the weights $\mathbf{W}$ of the first layer are encoded into the frequency domain. The output $z(t)$ of the mixer then contains the output pre-activations $\mathbf{z}$ of the first layer. The challenge then is to apply an elementwise nonlinearity on each frequency comb line of the output signal. A direct way of doing this is to pass $z(t)$ through an Analog-to-Digital Converter (ADC), perform a Fast Fourier Transform (FFT) to extract the amplitudes of the frequency comb lines $z_i$, apply the nonlinearity on them, perform an inverse FFT to obtain the time domain post-activation function $a(t)$, and then re-encode $a(t)$ into the analog domain through a Digital-to-Analog Converter (DAC) for subsequent multiplication with the weights of the next layer in the mixer. It turns out that the digital FFT operations in this approach can in fact be merged into the weight matrices and performed in analog with the help of IQ modulators; moreover, this `merging' trick enables us to encode weights and activations directly in the time-domain which in turn enables simple element-wise application of the nonlinearities on the time-domain encoded output activations (Fig.~\ref{fig:digitaltailored}). We present this digital-tailored protocol first. 

Next, we will move away entirely from attempting to faithfully mimic matrix-vector multiplications and move towards time-domain encoding of the weights and activations. The weights will be learned directly by training on the task loss function through the inherent mathematical behavior of the underlying analog hardware, including the nonidealities. In other words, the `digital twin' of the hardware will be used to train and obtain networks that are inherently optimized for the hardware's physics. We shall call this the `hardware-tailored' approach (see Fig.~\ref{fig:fullyanalog}). 

\subsubsection{Digital‑Tailored Pipeline}

The digital-tailored approach is demonstrated experimentally in our companion paper \cite{gao2025disaggregated} \textemdash\ we describe it here to lay the groundwork for the hardware-tailored approach. This pipeline merges FFT/IFFT into the weights via IQ modulation so that the client applies nonlinearities in the time domain without performing digital FFTs.

The composition of the $l$-th and $(l+1)$-th layers of the neural network can be represented by:
\begin{align}
    \mathbf{z}^{(l+1)}=\mathbf{W}^{(l+1)} \cdot \sigma\lp \mathbf{W}^{(l)}\mathbf{a}^{(l-1)}\rp.
    \label{eq:twolayer}
\end{align}
It should be kept in mind that all the above values are encoded in the frequency domain in MIWEN. We do this using single-sideband IQ modulators as discussed next. 

 \paragraph{IQ modulators}
The frequency encoded weight and activation signals in the MIWEN scheme can be expressed in the general language of single-sideband modulation as follows. Given an input sequence of $N$ complex numbers $a_1,a_2,\dots,a_N$, single-sideband modulation onto different frequency components produces the signal $a(t)$:
\begin{align}
    a(t)=\text{Re}\lp e^{i\omega_ct}\sum_ia_ie^{i\omega_it}\rp,
\end{align}
where $\omega_c$ is the carrier frequency. How would this be implemented in practice? Given a complex-valued envelope function $\tilde{a}(t)$, IQ modulators produce the signal $a(t)$:
\begin{align}
    a(t)=\text{Re}\lp e^{i\omega_ct}\tilde{a}(t)\rp
\end{align}
where $\omega_c$ is the carrier frequency. The envelope function $\tilde{a}(t)$ is assumed to be slowly-varying compared to the carrier $\omega_c$. This suggests that a way of encoding the MIWEN signals is to first Inverse Fourier Transform the $a_i$s into the envelope $\tilde{a}(t)$ and feed it into an IQ modulator to produce $a(t)$. We would like to avoid performing this ``encoding" Inverse Fourier Transform on the client's side. This can be done by lumping Fourier transform matrices $\mathbf{F}$ into Eq.~\eqref{eq:twolayer} as we show next.   

 \paragraph{Eliminating the encoding FFTs}
We introduce the $N\times N$ orthonormal Fourier transform matrix $\mathbf{F}_{N}$ and its conjugate transpose $\mathbf{F}^{\dagger}_{N}$ into Eq.~\eqref{eq:twolayer}:
\begin{align}
    \mathbf{z}^{(l+1)}&=\mathbf{W}^{(l+1)} \cdot \sigma\lp \mathbf{W}^{(l)}\mathbf{F}^{\dagger}_{N}\mathbf{F}_{N}\mathbf{a}^{(l-1)}\rp\\
    &=\mathbf{W}^{(l+1)} \cdot \sigma\lp \mathbf{W}^{(l)\text{pre}}\mathbf{a}^{\text{pre}(l-1)}\rp,
    \label{eq:eq13}
\end{align}
where we defined the pre-encoded activations $\mathbf{a}^{\text{pre}(l-1)}~:=~\mathbf{F}_{N}\mathbf{a}^{(l-1)}$ and pre-encoded weights $\mathbf{W}^{(l)\text{pre}}~:=~\mathbf{W}^{(l)}\mathbf{F}^{\dagger}_{N}$. We will now demand that $\mathbf{W}^{(l)\text{pre}}$ be broadcast by the server so that the client can process it with its local $\mathbf{a}^{\text{pre}(l-1)}$. From the previous section, we know that we simply have to perform an Inverse Fourier Transform on $\mathbf{a}^{\text{pre}(l-1)}$ to obtain samples of the its corresponding time domain $\tilde{a}^{\text{pre}}(t)$. But the Inverse Fourier Transform $\mathbf{F}^{\dagger}_{N}$ performed on $\mathbf{a}^{\text{pre}(l-1)}~=~\mathbf{F}_{N}\mathbf{a}^{(l-1)}$ simply yields $\mathbf{a}^{(l-1)}$! Therefore, the samples of $\tilde{a}^{\text{pre}}(t)$ that the IQ modulator needs in order to generate $a^{\text{pre}}(t)$ are simply the direct activations $\mathbf{a}^{(l-1)}$ themselves, with no need for any Fourier transform to be performed on it. This trick enables us to eliminate Fourier transforming the activations before encoding them using IQ modulators.   

 \paragraph{Eliminating the decoding FFTs}
We note that the output $\mathbf{W}^{(l)\text{pre}}\mathbf{a}^{\text{pre}(l-1)}$ in Eq.~\eqref{eq:eq13} is encoded in the frequency domain of the output waveform $z(t)$. Implementing the nonlinearity in the frequency domain is more difficult than doing it in the time domain. For this reason, we will actually demand that the weights sent by the server take on the following modified form:
\begin{align}
\mathbf{W}^{\text{post}(l)\text{pre}}~:=~\mathbf{F}_{M}\mathbf{W}^{(l)}\mathbf{F}^{\dagger}_{N}.
\end{align}
This ensures that the output recorded in the frequency domain of the output is $\mathbf{F}_{M}\mathbf{W}^{(l)}\mathbf{a}^{(l-1)}$, which implies that the time domain signal obtained when this output is sampled by an IQ demodulator is $\mathbf{W}^{(l)}\mathbf{a}^{(l-1)}$. The ReLU nonlinearity is then applied on the time bin output of the IQ demodulator via a simple diode before it is upconverted using another IQ modulator and passed onto a mixer for subsequent mixing with the incoming weights of the next layer $\mathbf{W}^{\text{post}(l+1)\text{pre}}$.    

 \paragraph{Filtration and nonlinearity in the time-domain}
The components of the vector $\mathbf{F}\mathbf{W}^{(l)}\mathbf{a}^{(l-1)}$ form a subset of the set of Fourier components of the output $z(t)$ of the mixer. The other components are spurious products that follow naturally from the frequency domain convolution performed by our protocol \cite{davis2022frequency}. $z(t)$ is downconverted by an IQ demodulator that yields the vector $\mathbf{W}^{(l)}\mathbf{a}^{(l-1)}$ as time-bin amplitudes as its output. The spurious frequencies are converted to time-bin pulses as well and are temporally filtered away through a switch. The neural network nonlinearity $\sigma(\cdot)$ is applied individually on the remaining time-bins before they are routed to the IQ modulator for mixing with the next wave of incoming weights from the server. This completes our description of the digital-tailored neural network protocol. A detailed, quantitative description of the protocol applied to a real system, together with its experimental demonstrations, is reported in our companion paper \cite{gao2025disaggregated}.

\subsubsection{Hardware‑Tailored Pipeline} \label{Hardware}
\begin{figure}[t]
\begin{centering}
\includegraphics[width=\columnwidth]{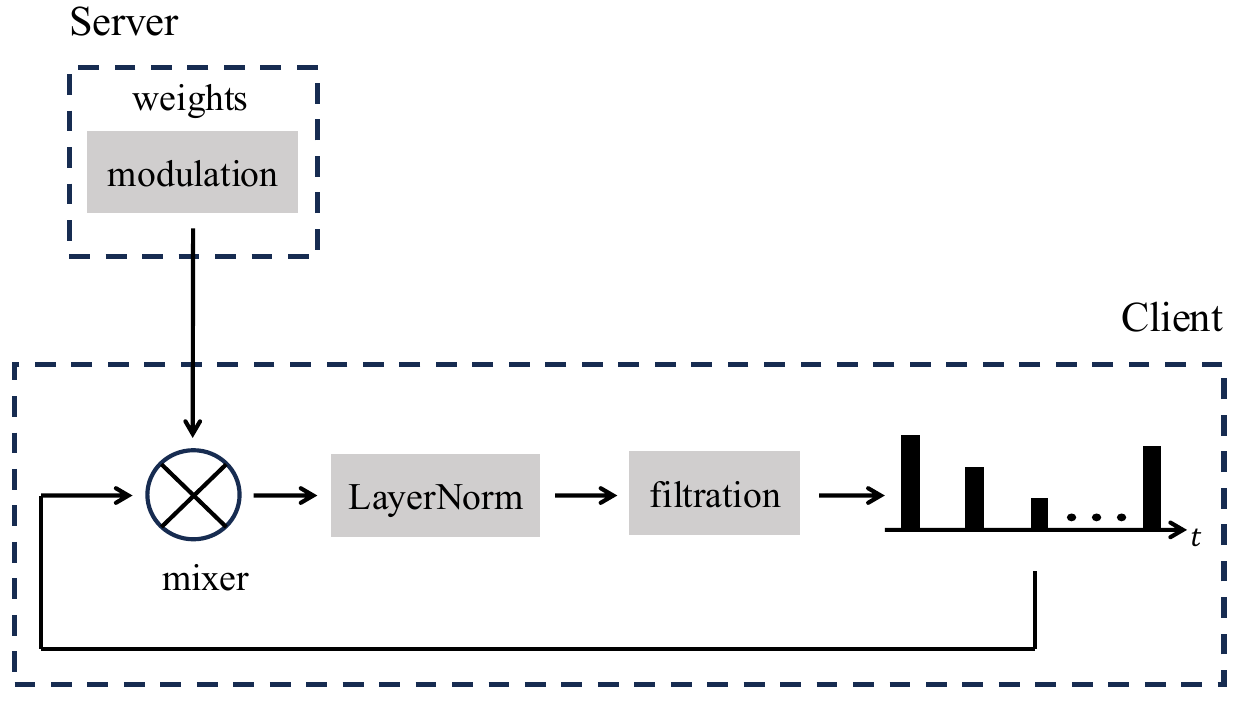}
\par\end{centering}
     \caption{\textbf{Hardware-tailored machine learning computation pipeline.} The server sends waveform $w(t)$ that gets mixed with the local time-encoded activations waveform $x(t)$ at the client's side. The result is sent through a layernorm layer (implemented via opamps) to stabilize the activations. The result is then passed through a linear filter to enable mixing between time-bins and enhance the expressivity of the physical neural network. The resultant final waveform is then mixed with the weights of the next layer.
     }
 \label{fig:fullyanalog}
 \end{figure}

In the hardware-tailored pipeline (Fig.~\ref{fig:fullyanalog}), weights and activations are encoded directly in time, mixed through a diode-ring nonlinearity, and then uses analog filtering and layer‑norm to provide inter‑time‑bin coupling and stability without any A/D interfaces between layers.
In the hardware‑tailored setting, we leverage the fact that our objective is not to compute exact inner products in any particular domain (time, frequency, or otherwise), but to enable energy-efficient inference that directly maps to our analog hardware. Therefore, we move away from reliance on precise numerical multiplication and accumulation (like we did earlier with frequency-encoding or time-encoding with FFT merging) and instead exploit the native physical properties of diode ring mixing and filtering to approximate the desired computation. 

To this end, we encode the weights and activations directly in the time domain in a fixed time window, using interpolation on the activations since they have fewer components than the weight matrix. These signals are passed through the diode ring mixer, which is time-instantaneous. Since the diode ring mixer does not allow for interaction between different time bins, we supplement the mixing process with filtration which mixes time bins to build a neural network layer. These layers are integrated into the training process, allowing the neural network to learn weights and representations that are inherently optimized for the hardware's physics.

The combination of analog mixing and filtering operations enables the system to perform expressive transformations of the input signals (Fig.~\ref{fig:fullyanalog}). Mixing shifts different input components to distinct frequencies, and filtering isolates or weights them selectively \textemdash\ together these operations provide sufficient computational flexibility for learning-based tasks like classification. By co-designing the physical layer operations and the network training procedure, we achieve high inference accuracy despite the lack of precise digital-style inner product computation. This approach eliminates the need for ADCs or digital logic between layers. All computation is performed directly on modulated analog signals, significantly reducing overhead while preserving task-specific performance.

 \paragraph{Stabilizing activations with a modified layer-norm}
For networks constructed with the mixing formula in Eq.~\eqref{eq:dioderingeq}, we noticed that training suffered from vanishing activations and gradients. To mitigate this problem, we introduced a modified, fully analog, layer-norm layer to control the activations and gradients \cite{ba2016layer} (Fig.~\ref{fig:fullyanalog}). 

Layer Normalization (LayerNorm) is a normalization technique that improves the training stability of deep neural networks by normalizing across the feature dimensions of each individual sample. Given a vector $\mathbf{x} = [x_1, x_2, \dots, x_d]$, LayerNorm computes the mean and variance as
\[
\mu = \frac{1}{d} \sum_{j=1}^d x_j, \quad \sigma^2 = \frac{1}{d} \sum_{j=1}^d (x_j - \mu)^2,
\]
and applies normalization followed by an affine transformation:
\[
\hat{x}_j = \frac{x_j - \mu}{\sqrt{\sigma^2 + \epsilon}}, \quad y_j = \gamma_j \hat{x}_j + \beta_j,
\]
where $\gamma_j$ and $\beta_j$ are learnable scaling and shifting parameters for each feature, and $\epsilon > 0$ ensures numerical stability.

We define a variant of LayerNorm that retains only the shifting component by fixing $\gamma_j = 1$ for all $j$. The resulting transformation becomes
\[
y_j = \hat{x}_j + \beta_j,
\]
preserving unit variance while allowing the mean to adapt through the learnable shift $\beta_j$. To enable further mixing between time-bins, we also include the same analog bandpass filters after each layernorm module that were introduced in Section~\ref{sec2.1} in the context of Eq.~\eqref{eq:matvectprod}. In summary, each layer of the hardware-tailored network consists of diode ring mixing followed by layernorm and filtration (Fig.~\ref{fig:fullyanalog}).

\subsubsection{Numerical Results}
To evaluate the achievable test accuracy as a function of energy cost, we wrote custom layers in the popular machine learning package PyTorch that describe the operation of each step of our protocol. This includes all the components of the computation chain: encoding of activation vectors and weights into signals $x(t),w(t)$, incorporation of thermal noise at the appropriate points, and a physically accurate description of the diode ring mixing process. The custom layers were then composed into neural networks that were trained on the MNIST (Modified National Institute of Standards and Technology) digit recognition task.

\begin{figure}[t]
\begin{centering}
\includegraphics[width=\columnwidth]{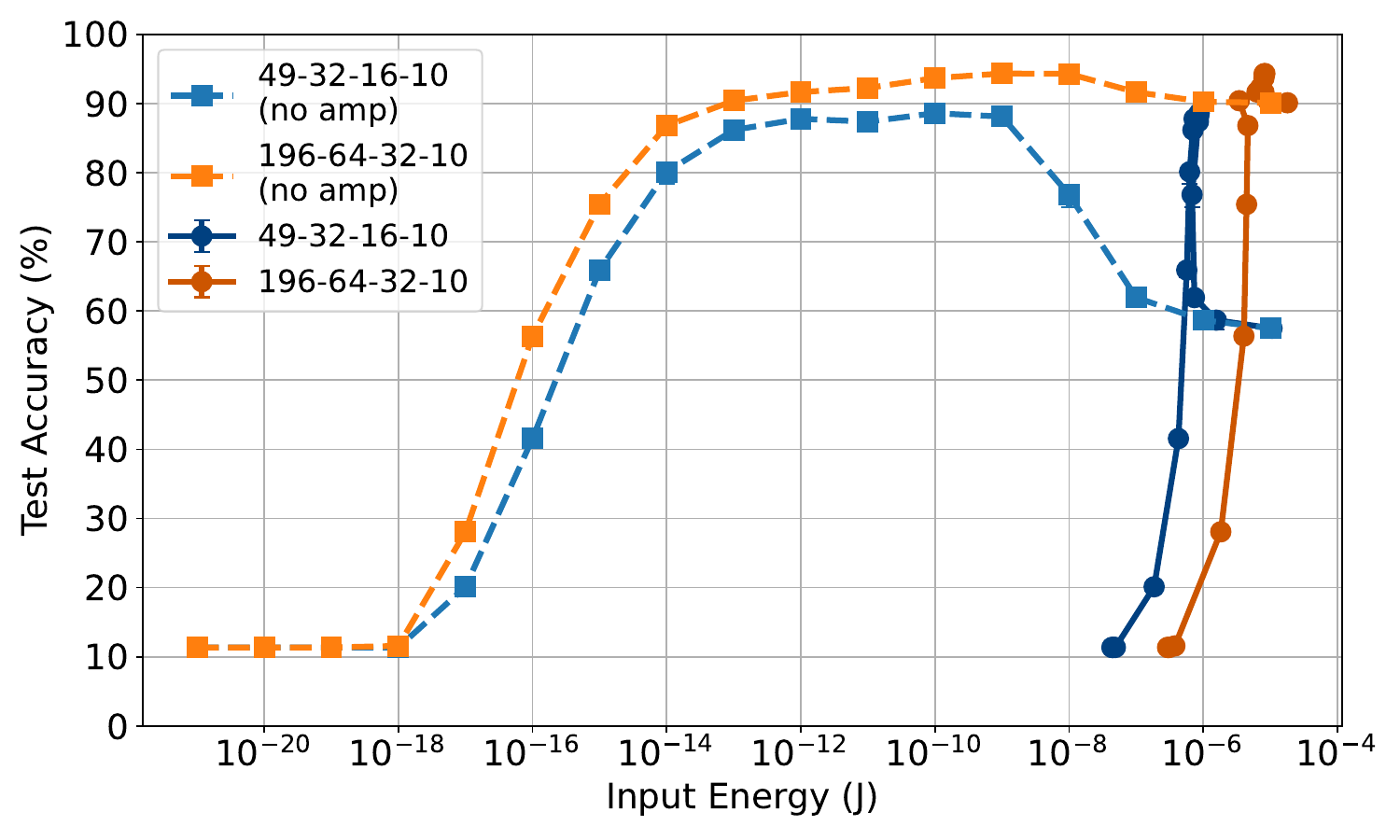}
\par\end{centering}
     \caption{\textbf{MNIST classification accuracy of RF analog deep learning.} MNIST classification accuracy is presented as a function of the total energy per inference for MIWEN. 
    The legend indicates the network architecture and whether the energy of the layernorm implicit amplification is taken into account. The test accuracy starts out at random guessing, increases to near-digital accuracy, and then deteriorates again as the diode mixer deviates from exact multiplication $w(t)x(t)$. The dotted lines depict the test accuracy as a function of the energy of the client input waveform $x(t)$, while the solid lines depict the same test accuracies but now include the amplification energy implicitly expended in the layernorm layers. Accuracy reaches 95\% on MNIST at 100\,pJ input energy before degrading at the nJ level as the diode nonlinearity departs from ideal multiplication. Results are for 5 models trained at each energy level; the median and inter-quartile range are presented as the center point and error bars respectively.
     }
 \label{fig:results}
 \end{figure}

Fig.~\ref{fig:results} presents the variation in test accuracy as a function of client energy for two different network architectures (smaller: 49-32-16-10 and larger: 196-64-32-10) across two energy-accounting modes: activation energy and total energy. The energy thresholds are applied to the client-side processing. Square markers represent test accuracy versus activation energy alone (the energy associated with the client input waveform \(x(t)\)), while circle markers account for total energy, which includes both the activation energy and the implicit amplification introduced by the LayerNorm modules (see Sec.~\ref{Hardware}). As noted earlier, LayerNorm enhances training stability and facilitates convergence. However, recent research \cite{zhu2025transformers} has demonstrated the potential for normalization-free training, offering a promising direction for reducing the amplification overhead in MIWEN. The markers represent the median, and the error bars represent the interquartile range. For each energy point, the weights and biases, broadcast by the server, are optimized to yield the corresponding classification accuracy. Since our system operates in a disaggregated memory configuration—where the weights \(w(t)\) are streamed from a remote server—we focus solely on the client-side energy required to process these weights, referred to as `input energy' in Fig.~\ref{fig:results}. To generate inputs of the correct size, MNIST images were downsampled by patch-averaging from their original 28 × 28 resolution.

The performance of both networks as a function of the activation energy (square markers) is equivalent to random guessing below the attojoule (aJ) level. As energy increases, accuracy rises sharply, reaching near-digital performance of about 95\% at $100$\,pJ. Beyond the nanojoule (nJ) regime, however, the diode ring mixer departs from ideal multiplicative behavior, and accuracy degrades, with the severity depending on network size. The accuracy as a function of the total energy rises from $100$\,nJ, the typical energy required for the Layernorm amplification, and achieves near digital accuracy around microjoule level $\mu$J. These results show that the same RF chain used for communication in edge devices can also carry out machine-learning inference with accuracy close to digital implementations, when operated within the typical energy constraints of standard front-end receivers in such devices.

\section{Discussion}
We introduce Machine Intelligence on Wireless Edge Networks (MIWEN), an approach that reuses the RF receive chain already embedded in billions of edge devices to perform neural inference. Instead of storing the model locally or offloading queries to a server, MIWEN broadcasts model weights as RF waveforms and computes multiply–accumulate operations directly in the client’s mixer and filter stages. This design eliminates the need for local weight storage and repeated signal conversions, while building entirely on hardware that is already present in commodity transceivers. 

Our results demonstrate that inference accuracy improves with client input energy and reaches near-digital levels within the standard energy budgets of wireless communication front ends in edge devices. These findings confirm that the same RF hardware already used for communication can also execute machine-learning inference without additional hardware overhead. By incorporating a differentiable model of the RF chain into training, our hardware-tailored pipelines maintain accuracy without requiring conversions between layers. 

Looking ahead, MIWEN opens opportunities for large-scale deployments involving multiple clients and servers, where scheduling becomes critical to balance computation and communication. Further work may explore optimizations for network constraints such as bandwidth, latency and power allocation, and evaluate the approach under realistic traffic conditions. Together, these directions will help establish MIWEN as a practical framework for distributed, low-power intelligence at the wireless edge.

\section{Methods}

\subsection{Deriving the Effective Number of Bits (ENOB) for Analog Computation} \label{met_4.1}
\subsubsection{Noisy Analog Inner‑Product Engine}
Let $\mathbf{w},\mathbf{x}$ be vectors from $\mathbb{R}^d$ whose components are individually independently drawn from arbitrary mean-zero probability distributions. Consider their digital inner product
\[
y_{\mathrm{digital}} \;=\; \mathbf{w}\,\cdot\,\mathbf{x},
\]
and the noisy analog counterpart
\[
y_{\mathrm{analog}} \;=\; \mathbf{w}\,\cdot\,\mathbf{x} \;+\; n,
\]
where \(n\) is additive Gaussian noise, $n \;\sim\; \mathcal{N}(0,\sigma_n^2)$.

We wish to quantify how many bits of information about \(y_{\mathrm{digital}}\) are contained in \(y_{\mathrm{analog}}\).  By definition, the mutual information between \(y_{\mathrm{digital}}\) and \(y_{\mathrm{analog}}\) is
\[
I\bigl(y_{\mathrm{digital}};\,y_{\mathrm{analog}}\bigr)
\;=\;
H\bigl(y_{\mathrm{analog}}\bigr)
\;-\;
H\bigl(y_{\mathrm{analog}}\,\bigm|\,y_{\mathrm{digital}}\bigr).
\]
Because once \(y_{\mathrm{digital}} = \mathbf{w}\cdot\mathbf{x}\) is known, the only remaining uncertainty in \(y_{\mathrm{optical}}\) is the Gaussian noise \(n\), we have
\[
H\bigl(y_{\mathrm{analog}} \,\bigm|\, y_{\mathrm{digital}}\bigr)
\;=\;
H(n)
\;=\;
\frac{1}{2}\,\log_{2}\bigl(2\pi e\,\sigma_n^2\bigr).
\]

To evaluate \(H\bigl(y_{\mathrm{analog}}\bigr)\), we note that \(\mathbf{w}\cdot\mathbf{x}\) is a sum of (not necessarily identically distributed) independent random variables and appeal to the Central Limit Theorem to obtain,
\[
\mathbf{w}\,\cdot\,\mathbf{x}
\;\sim\;
\mathcal{N}\bigl(0,\;\mathbb{E}[(\mathbf{w}\cdot\mathbf{x})^2]\bigr),
\]
so that
\begin{align}
y_{\mathrm{analog}}
\;=&\;
\underbrace{\mathbf{w}\,\cdot\,\mathbf{x}}_{\mathcal{N}(0,\;\mathbb{E}[(\mathbf{w}\cdot\mathbf{x})^2])}
\;+\;
\underbrace{n}_{\mathcal{N}(0,\;\sigma_n^2)}
\;\\
&\sim\;
\mathcal{N}\Bigl(0,\;\mathbb{E}[(\mathbf{w}\cdot\mathbf{x})^2] \;+\; \sigma_n^2\Bigr).
\end{align}
Hence its differential entropy is
\[
H\bigl(y_{\mathrm{analog}}\bigr)
\;=\;
\frac{1}{2}\,\log_{2}\!\Bigl(2\pi e\,\bigl[\mathbb{E}[(\mathbf{w}\cdot\mathbf{x})^2] + \sigma_n^2\bigr]\Bigr).
\]
Putting these pieces together gives
\[
I\bigl(y_{\mathrm{digital}};\,y_{\mathrm{analog}}\bigr)
\;=\;
\frac{1}{2}\,\log_{2}\!\Bigl(1 \;+\; \frac{\mathbb{E}[(\mathbf{w}\cdot\mathbf{x})^2]}{\sigma_n^2}\Bigr).
\]
By defining the signal‐to‐noise ratio \(\mathrm{SNR} = \frac{\mathbb{E}[(\mathbf{w}\cdot\mathbf{x})^2]}{\sigma_n^2}\), we obtain the result
\[
I\bigl(y_{\mathrm{digital}};\,y_{\mathrm{analog}}\bigr)
\;=\;
\frac{1}{2}\,\log_{2}\bigl(1 + \mathrm{SNR}\bigr).
\]
Finally, identifying this mutual information (in bits) with the effective number of bits (ENOB) of the analog computation yields
\[
\mathrm{ENOB}
\;=\;
I\bigl(y_{\mathrm{digital}};\,y_{\mathrm{analog}}\bigr)
\;=\;
\frac{1}{2}\,\log_{2}\bigl(1 + \mathrm{SNR}\bigr).
\]

\subsubsection{Noisy Analog Diode‑Ring Mixer}
We note in this section that we do not have an exact analog inner product engine in our system; instead our system performs
\begin{align}
y_{\mathrm{analog}} \;=\; f(\mathbf{w}+n_w,\mathbf{x}+n_x) \;+\; n    
\end{align}
where \(n, n_w, n_x\) are Gaussian noise random variables. To fit this into the model we discussed in the previous subsection, we rewrite his equation as:
\begin{align}
y_{\mathrm{analog}} \;&=\; \mathbf{w}\,\cdot\,\mathbf{x}+\left(f(\mathbf{w}+n_w,\mathbf{x}+n_x)-\mathbf{w}\,\cdot\,\mathbf{x} \;+\; n\right)\\
&\approx\mathbf{w}\,\cdot\,\mathbf{x}+n'\notag
\end{align}
where $n'$ captures the total discrepancy between the true digital inner product and the analog inner product due to both noise and functional differences. To compute the ENOB, we will continue to use the formula
\[
I\bigl(y_{\mathrm{digital}};\,y_{\mathrm{analog}}\bigr)
\;=\;
\frac{1}{2}\,\log_{2}\bigl(1 + \mathrm{SNR}\bigr).
\]
with the same signal definition $S=\mathbb{E}\left[\left(\mathbf{w}\cdot\mathbf{x}\right)^2\right]$ but with the noise now defined as $N=\mathbb{E}[n'^2]$. We used this definition to compute the variation of the inner product ENOB with client and server power in Fig. 2 in the main text. 

\subsection{SNR and Energy per MAC of the RF Inner‑Product Engine}\label{app:SNR}

\subsubsection{SNR Estimation}
We shall estimate the SNR of the RF mixer by working at low voltages where the noisy RF mixing process can be approximated by:
\begin{align*}
    y_{\text{RF}}(t) = \frac{e}{4kT} \cdot \left[w(t)+n_w(t)\right] \cdot \left[x(t)+n_x(t)\right] + n_y(t).
\end{align*}
In our implementation, the activations $x_i$ and the weights $w_i$ are encoded into orthonormal frequency basis, thus yielding the following noisy inner product model:
\begin{align}
    y=\frac{e}{4kT}\frac{1}{\sqrt{N}}\ls\sum_i\lp w_i+n^{(w)}_i\rp\lp x_i+n^{(x)}_i\rp\rs+n^{(y)}
\end{align}
The components $w_i$ and $x_j$ are drawn independently from normal distributions as mentioned in the main text. The noise components $n^{(w)}_i,n^{(x)}_j,n^{(y)}$ are all assumed to be independently drawn from normal distributions of mean $0$ and variance $4kTR\Delta f$. 

We note that we have a peculiar situation where the signal is produced by an underlying random process and the noise on top of it is also produced by a different underlying random process. We define the SNR for this kind of situation in the following way. We first compute the squared average output $\la y\ra_n^2$ for a given $w,x$, but averaged over all the noise components:
\begin{align}
    \la y\ra_n^2=\lp\frac{e}{4kT}\rp^2\frac{1}{N}\lp\sum_iw_ix_i\rp^2=\lp\frac{e}{4kT}\rp^2\frac{1}{N}\lp\mathbf w\cdot\mathbf x\rp^2
\end{align}
The quantity above will be treated as the ``signal strength" for the given $w,x$. The ``noise strength" for the same $w,x$ is given by the variance of $y$ computed over the noise random variables:

\begin{align}
    \text{Var}_n(y) &= \left(\frac{e}{4kT}\right)^2 \frac{1}{N} \sum_i \text{Var}_n \left( \left( w_i + n^{(w)}_i \right) \left( x_i + n^{(x)}_i \right) \right) \notag \\
    &\quad + \text{Var}_n \left( n^{(y)} \right) \\
    &= \left( \frac{e}{4kT} \right)^2 \frac{1}{N} \sum_i \left( \sigma_w^2 x_i^2 + \sigma_x^2 w_i^2 + \sigma_w^2 \sigma_x^2 \right) + \sigma_y^2 \notag
\end{align}

Now, we shall define the signal and noise strengths for our overall setup as the average of $\la y\ra_n^2$ and $\text{Var}_n(y)$ over all choices of $w,x$. For this purpose, we assume that the $w_i, x_i$ have variances $M_w^2, M_x^2$ respectively:
\begin{align}
    S&=\la\la y\ra_n^2\ra_{w,x}=\lp\frac{e}{4kT}\rp^2M_w^2M_x^2d/N\\
    \text{Noise}&=\la\text{Var}_n(y)\ra_{w,x}\\
    &=\lp\frac{e}{4kT}\rp^2\frac{d}{N}\lp M_x^2\sigma^2_w+M_w^2\sigma^2_x+\sigma^2_w\sigma^2_x\rp+\sigma^2_y
\end{align}
The SNR of our inner product engine is then:
\begin{align}
    \text{SNR}=\frac{S}{\text{N}}=\frac{\lp\frac{e}{4kT}\rp^2M_w^2M_x^2d/N}{\lp\frac{e}{4kT}\rp^2\frac{d}{N}\lp M_x^2\sigma^2_w+M_w^2\sigma^2_x+\sigma^2_w\sigma^2_x\rp+\sigma^2_y}
\end{align}
For encoding in the orthonormal Fourier domain, we have $\sigma_w^2=\sigma_x^2=\sigma_y^2=4kTR\Delta f$, where we chose to keep the total bandwidth constant even when the input size is varied. Since $d$ and $N$ are essentially proportional to one another, the formula predicts that the SNR is independent of $d$ for $d\gtrsim10$.

\subsubsection{Energy per MAC}
The total energy $E_x$ of the $x(t)$ waveform generated by the client is:
\begin{align}
    E_x=\int dt \frac{x^2(t)}{R}\approx\Delta t\sum_n\frac{x^2[n]}{R} 
\end{align}
Since we encode our inputs in the orthonormalized Fourier domain, Parseval's theorem states that $\sum_n|x[n]|^2=\sum_k|X[k]|^2$. Since the $d$ $X[k]$s that correspond to the activation vector each have variance $M_x^2$, the client energy per MAC is:
\begin{align}
    \frac{E_x}{d}\approx\frac{\Delta t}{d}\sum_k\frac{|X[k]|^2}{R}\approx \Delta t\frac{M_x^2}{R}
\end{align}
The signal energy per MAC is:
\begin{align}
    \frac{E_w}{d}\approx\frac{\Delta t}{d}\sum_k\frac{|W[k]|^2}{R}\approx \Delta t\frac{M_w^2}{R}
\end{align}


\subsection{Circuit Analysis} \label{Met_cir}
\begin{figure}[ht!]
\begin{centering}
\includegraphics[scale=0.32]{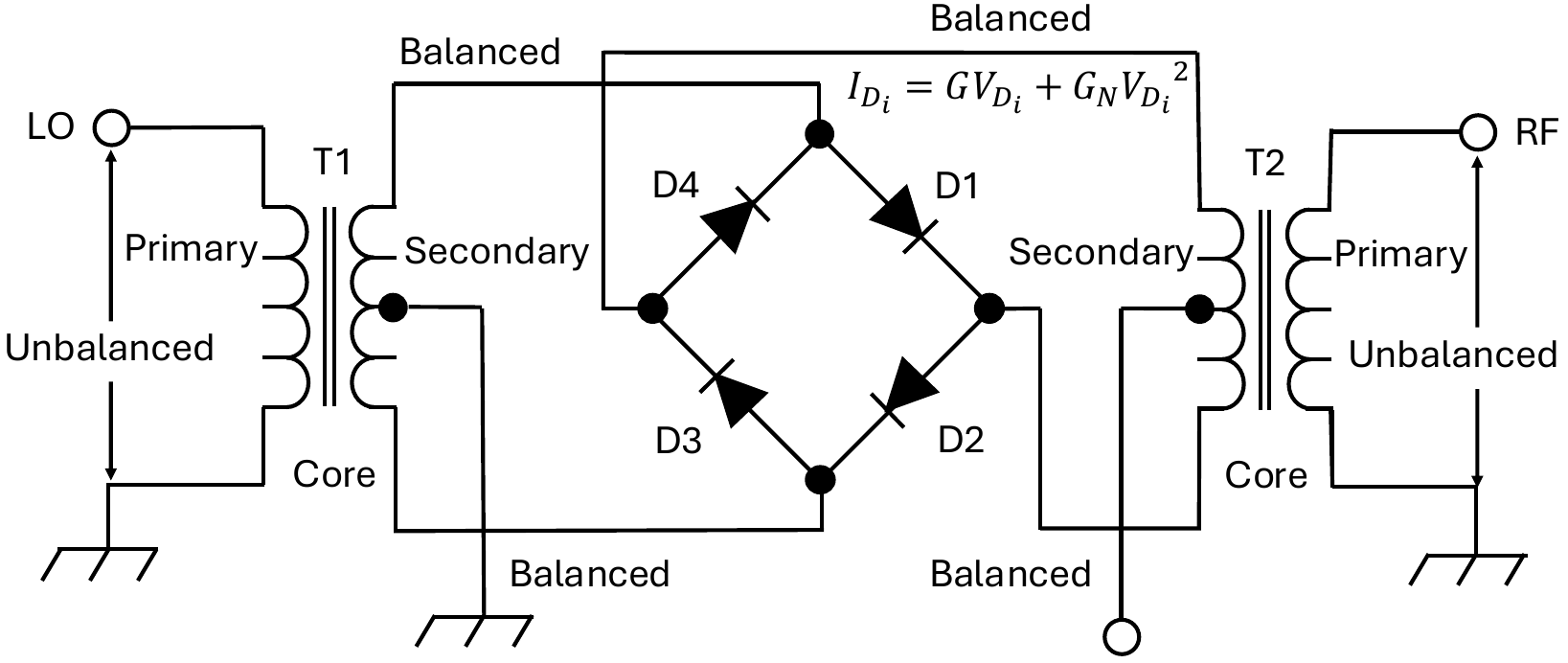}
\par\end{centering}
\caption{Schematic of the passive diode ring mixer. The diagram illustrates the signal path through the transformers (T1 and T2), the core mixer circuit, and the four diodes (D1, D2, D3, D4) involved in the frequency mixing process. The current \( I_D \) through each diode is related to the voltage \( V_D \) by the expression \( I_D = G V_D + G_N V_D^2 \), where \( G \) and \( G_N \) are the conductance parameters.}
 \label{circuit}
 \end{figure}

Here we utilize a passive diode ring mixer to perform the signal mixing in the radio-frequency (RF) domain. The schematic of the diode ring mixer is shown in Fig.~\ref{circuit}, where the signal path is routed through balanced and unbalanced transformer stages (T1 and T2), with the core mixer circuit composed of four diodes (D1, D2, D3, D4). The current \( I_D \) flowing through each diode is given by the equation \( I_D = G V_D + G_N V_D^2 \), where \( G \) and \( G_N \) represent the conductance parameters, and \( V_D \) is the diode voltage. This setup enables the core functionality of the mixer while minimizing energy losses associated with unnecessary signal conversions.

\subsubsection{Diode‑Ring Small‑Signal Derivation} 
 We first write out the characteristic equation of the diode, and then expand it up to the second order in Taylor series.
 \begin{align}
     I&=I_0\lp e^{\frac{qV}{kT}}-1\rp=I_0\lp e^{\frac{V}{V_T}}-1\rp\\
     &=I_0\frac{V}{V_T}+I_0\frac{V^2}{2V_T^2}=GV+G_NV^2
 \end{align}
where we set $V_T:=kT/q$. Denoting the potentials at the four corners of the ring by $A, B, C, D$, and writing out the voltage equations implied by the circuit connectivity, we get:
\begin{align}
    A-C&=V_{\text{LO}}\\
    A+C&=0\\
    D-B&=V_{\text{RF}}\\
    I_A&=I_C,\ I_B=I_D\\
    V_{\text{IF}}&=(B+D)/2
\end{align}
The current going into $A$ is:
 \begin{align}
     I_A =&G\lp n_1+n_4+V_{\text{LO}}-2B-V_{\text{RF}}\rp\\
     &+G_N\lp -V_{\text{RF}}^2-2BV_{\text{RF}}+V_{\text{RF}}V_{\text{LO}}\rp
 \end{align}
 \begin{align}
     I_A =G\lp V_{\text{LO}}-2B-V_{\text{RF}}\rp+\\ G_N\lp -V_{\text{RF}}^2-2BV_{\text{RF}}+V_{\text{RF}}V_{\text{LO}}\rp
 \end{align}
 while the current leaving $C$ is:
 \begin{align}
     I_C =&G\lp n_2+n_3+ V_{\text{LO}}+2B+V_{\text{RF}}\rp\\
     &+G_N\lp -V_{\text{RF}}^2-2BV_{\text{RF}}-V_{\text{RF}}V_{\text{LO}}\rp
 \end{align}
 \begin{align}
     I_C =G\lp V_{\text{LO}}+2B+V_{\text{RF}}\rp\\+G_N\lp -V_{\text{RF}}^2-2BV_{\text{RF}}-V_{\text{RF}}V_{\text{LO}}\rp
 \end{align}
Setting $I_A=I_C$, we get:
\begin{align}
    B&=-\frac{V_{\text{RF}}}{2}+\frac{G_N}{2G}V_{\text{RF}}V_{\text{LO}}\\
    D&=\frac{V_{\text{RF}}}{2}+\frac{G_N}{2G}V_{\text{RF}}V_{\text{LO}}
\end{align}
The output we are interested in, $V_{\text{IF}}$, is then:
\begin{align}
    V_{\text{IF}}&=\frac{G_N}{2G}V_{\text{RF}}V_{\text{LO}}+\frac{n_1+n_4-n_2-n_3}{4}\\
    &=\frac{1}{4V_T}V_{\text{RF}}V_{\text{LO}}+\frac{n_1+n_4-n_2-n_3}{4}
\end{align}
\begin{align}
    V_{\text{IF}}=\frac{G_N}{2G}V_{\text{RF}}V_{\text{LO}}=\frac{1}{4V_T}V_{\text{RF}}V_{\text{LO}}\approx 10V_{\text{RF}}V_{\text{LO}}
\end{align}

\subsubsection{Diode‑Ring Exact Derivation}
The diode's characteristic:
\begin{align}
     I&=I_0\lp e^{\frac{qV}{kT}}-1\rp=I_0\lp e^{\frac{V}{V_T}}-1\rp
\end{align}
where we set $V_T:=kT/q$. Denoting the potentials at the four corners of the ring by $A, B, C, D$, and writing out the voltage equations implied by the circuit connectivity, we get:
\begin{align}
    A-C&=V_{\text{LO}}\\
    A+C&=0\\
    D-B&=V_{\text{RF}}\\
    I_A&=I_C,\ I_B=I_D\\
    V_{\text{IF}}&=(B+D)/2
\end{align}
The current going into $A$ is:
 \begin{align}
     I_A =I_0\lp e^{\frac{V_{lo}/2-B}{V_T}}-e^{\frac{B+V_{rf}-V_{lo}/2}{V_T}}\rp
 \end{align}
 while the current leaving $C$ is:
 \begin{align}
     I_C =I_0\lp e^{\frac{V_{lo}/2+B}{V_T}}-e^{\frac{-B-V_{rf}-V_{lo}/2}{V_T}}\rp
 \end{align}
Setting $I_A=I_C$, we get:
\begin{align}
    B&=\frac{V_T}{2}\ln{\lp\frac{e^{\frac{V_{lo}}{2V_T}}+e^{\frac{-V_{rf}-V_{lo}/2}{V_T}}}{e^{\frac{V_{lo}}{2V_T}}+e^{\frac{V_{rf}-V_{lo}/2}{V_T}}}\rp}
\end{align}
Then, the output is:
\begin{align}
V_{if}=\frac{V_{rf}}{2}+\frac{V_T}{2}\ln{\lp\frac{e^{\frac{V_{lo}}{V_T}}+e^{\frac{-V_{rf}}{V_T}}}{e^{\frac{V_{lo}}{V_T}}+e^{\frac{V_{rf}}{V_T}}}\rp}
\end{align}

\subsubsection{Second‑Order Expansion Derivation}

We want to show that
\[
z \;=\; \frac{y}{2} \;+\; \frac{VT}{2}\,\ln\!\Bigl(\tfrac{e^{x/VT}+e^{-y/VT}}{e^{x/VT}+e^{y/VT}}\Bigr)
\]
reduces, up to second order in \(\frac{x}{VT}\) and \(\frac{y}{VT}\), to
\[
z \;=\; \frac{x\,y}{4\,VT}.
\]

\paragraph{Introduce dimensionless parameters}

Define
\[
\alpha \;=\; \frac{x}{VT}, 
\quad
\beta \;=\; \frac{y}{VT}.
\]
Then
\[
z 
\;=\;
\frac{y}{2}
\;+\;
\frac{VT}{2}
\;\ln\!\Bigl(
  \frac{e^\alpha + e^{-\beta}}{\,e^\alpha + e^\beta\,}
\Bigr).
\]
We focus on the argument of the logarithm:
\[
\ln\!\Bigl(
  \frac{e^\alpha + e^{-\beta}}{e^\alpha + e^\beta}
\Bigr)
\;=\;
\ln\!\bigl(\,N/D\bigr),
\]
where
\[
N \;=\; e^\alpha + e^{-\beta},
\quad
D \;=\; e^\alpha + e^\beta.
\]

\paragraph{Expand exponentials to second order}

For small \(\alpha\) and \(\beta\), we use the expansions:
\[
\begin{aligned}
e^\alpha &\approx 1 + \alpha + \tfrac12\,\alpha^2,\\
e^\beta  &\approx 1 + \beta  + \tfrac12\,\beta^2,\\
e^{-\beta} &\approx 1 - \beta + \tfrac12\,\beta^2.
\end{aligned}
\]
Hence, up to second order,
\[
\begin{aligned}
N &= e^\alpha + e^{-\beta}
\;\approx\;
\bigl(1 + \alpha + \tfrac12\,\alpha^2\bigr)
\;+\;
\bigl(1 - \beta + \tfrac12\,\beta^2\bigr)
\\[6pt]
&= 2 + (\alpha - \beta) + \tfrac12(\alpha^2 + \beta^2),
\\[8pt]
D &= e^\alpha + e^\beta
\;\approx\;
\bigl(1 + \alpha + \tfrac12\,\alpha^2\bigr)
\;+\;
\bigl(1 + \beta + \tfrac12\,\beta^2\bigr)
\\[6pt]
&= 2 + (\alpha + \beta) + \tfrac12(\alpha^2 + \beta^2).
\end{aligned}
\]
For compactness, set
\[
A \;=\; \alpha + \beta,
\quad
B \;=\; \alpha - \beta,
\quad
Q \;=\; \tfrac12\,(\alpha^2 + \beta^2).
\]
Then
\[
N = 2 + B + Q,
\quad
D = 2 + A + Q.
\]
We want the ratio
\[
R \;=\; \frac{N}{D}
\;=\;
\frac{2 + B + Q}{\,2 + A + Q\,}
\]
only up to second order in \(\alpha\) and \(\beta\).

\paragraph{Expand \(\frac{N}{D}\) to second order}

Factor out 2:
\[
R 
\;=\; 
\frac{2 \bigl(1 + \tfrac{B+Q}{2}\bigr)}{\,2 \bigl(1 + \tfrac{A+Q}{2}\bigr)}
\;=\;
\frac{1 + X}{\,1 + Y},
\]
where
\[
X = \frac{B + Q}{2},
\quad
Y = \frac{A + Q}{2}.
\]
For small \(X, Y\), we expand:
\[
\frac{1 + X}{1 + Y}
\;=\;
1 + (X - Y) - \frac{(X - Y)^2}{2} + \cdots,
\]
but it is sufficient just to keep track of terms through second order.  After collecting terms carefully, one finds
\[
R 
\;=\;
1 + \bigl(-\beta + \tfrac{\alpha\,\beta}{2} + \tfrac{\beta^2}{2}\bigr) 
+ \cdots
\;=\;
1 + r,
\]
where
\[
r
\;=\;
-\,\beta 
\;+\; 
\frac{\alpha\,\beta}{2}
\;+\;
\frac{\beta^2}{2}
\;+\;\cdots.
\]

\paragraph{Expand \(\ln(R)\) to second order}

Since \(R = 1 + r\) with \(r\) small,
\[
\ln(R)
\;=\;
\ln(1 + r)
\;=\;
r - \frac{r^2}{2} + \cdots.
\]
Here \(r\) itself is already first- plus second-order, so
\[
r^2 \approx \beta^2
\quad (\text{since the first-order part of }r\text{ is }-\beta).
\]
Thus,
\[
\ln(R)
\;=\;
\Bigl(-\beta + \frac{\alpha\,\beta}{2} + \frac{\beta^2}{2}\Bigr)
\;-\;
\frac{\beta^2}{2}
\;+\;\cdots
\;=\;
-\,\beta 
\;+\;
\frac{\alpha\,\beta}{2}
\;+\;\cdots.
\]
Recalling \(\beta = \frac{y}{VT}\) and \(\alpha = \frac{x}{VT}\):
\[
\ln(R)
\;\approx\;
-\frac{y}{VT}
\;+\;
\frac{x\,y}{2\,(VT)^2}.
\]

\paragraph{Substitute back into \(z\)}

We have
\[
z 
\;=\;
\frac{y}{2}
\;+\;
\frac{VT}{2}
\;\ln(R).
\]
Therefore,
\[
\frac{VT}{2}\,\ln(R)
\;=\;
\frac{VT}{2}
\bigl(
-\tfrac{y}{VT} + \tfrac{x\,y}{2\,(VT)^2}
\bigr)
\;=\;
-\frac{y}{2} 
\;+\;
\frac{x\,y}{4\,VT}.
\]
Hence
\[
z 
\;=\;
\underbrace{\frac{y}{2}}_{\text{first part}}
\;+\;
\underbrace{\Bigl(-\tfrac{y}{2} + \tfrac{x\,y}{4\,VT}\Bigr)}_{\frac{VT}{2}\,\ln(R)}
\;=\;
\frac{x\,y}{4\,VT}.
\]
This completes the derivation.


\section*{Acknowledgments}
S.K.V. was supported by the NSF RAISE-TAQS program, grant number 1936314, and the NTT Research Inc. grants ``Large-scale nanophotonic circuits for neuromorphic computing'' and ``Netcast'', administered by MIT. K.S. acknowledges the support of the Israeli Council for Higher Education and the Zuckerman STEM Leadership Program. D.E. acknowledges partial support from the NSF RAISE-TAQS program (grant number 1936314), NSF C-Accel program (grant number 2040695), and DARPA QuANET program. Z.G. and T.C. acknowledge partial support from the NSF Athena AI Institute for Edge Computing (CNS-2112562). 

\section*{Author Contributions}
S.K.V. and K.S performed the analysis and simulations under the supervision of D.E. Z.G. and T.C. contributed to the design and methods. All authors contributed
to the writing.

\section*{Data and Code Availability}
All the data and code for this paper have been made available at the following GitHub link: \href{https://github.com/QPG-MIT/MIWEN}{https://github.com/QPG-MIT/MIWEN}

\bibliographystyle{unsrtnat} 
\bibliography{refs}          



\end{document}